# Unitary Quantum Lattice Simulations for Maxwell Equations in Vacuum and in Dielectric Media

George Vahala, Department of Physics, William & Mary, Williamsburg, VA 23185
Linda Vahala, Department of Electrical & Computer Enigneering, Old Dominion University, Norfolk, VA 23529
Min Soe, Department of Mathematics and Physical Sciences, Rogers State University, Claremore, OK 74017
Abhay K. Ram, Plasma Science and Fusion Center, MIT, Cambridge, MA 02139

**Abstract**
Utilizing the similarity between the spinor representation of the Dirac equation and the Maxwell equations that has been recognized since the early days of relativistic quantum mechanics, a quantum lattice (QLA) representation of unitary collision-stream operators of Maxwell's equations is derived for both homogeneous and inhomogeneous media. A second order accurate 4-spinor scheme is developed and tested successfully for two dimensional (2D) propagation of a Gaussian pulse in a uniform medium while for normal (1D) incidence of an electromagnetic Gaussian pulse onto a dielectric interface requires 8-component spinors. In particular, the well-known phase change, field amplitudes and profile widths are recovered by the QLA asymptotic profiles without the imposition of electromagnetic boundary conditions at the interface. The QLA simulations yield the time-dependent electromagnetic fields as the pulse enters and straddles the dielectric boundary. QLA involves unitary interleaved non-commuting collision and streaming operators that can be coded onto a quantum computer – the non-commutation being the only reason why one perturbatively recovers the Maxwell equations.

## I. INTRODUCTION

Dirac [1] derived a relativistic covariant representation of the Schrodinger equation with positive definite probability density by, in essence, taking the square root of the Klein-Gordon wave equation. With the introduction of Dirac spinors, there were immediate attempts to connect Maxwell's equations with the Dirac equation [2, 3], particularly with the introduction of the Riemann-Silberstein vector [4] for the electromagnetic field. More recent attempts have also coupled the Maxwell equations to various field theories [5-7].

Here we will give an explicit unitary quantum lattice algorithm (QLA) for Maxwell equations in material media, building on our earlier QLA for solitons [8-11, 17-19] and Bose-Einstein condensates [12-16, 20-23]. QLA are of much interest since its interleaved sequence of unitary collision and streaming operators can be immediately modeled by qubit gates thereby permitting immediate encoding onto a quantum computer. An interesting by-product of QLA is that these algorithms are also ideally parallelizable on classical supercomputers and can lead to algorithms that can outperform standard classical algorithms.

Khan [24] has expressed the Maxwell equations

$$\nabla \cdot \mathbf{D}(\mathbf{x},t) = \rho(\mathbf{x},t) \quad , \quad \nabla \cdot \mathbf{B}(\mathbf{x},t) = 0$$
$$\nabla \times \mathbf{H}(\mathbf{x},t) = \mathbf{J}(\mathbf{x},t) + \frac{\partial \mathbf{D}(\mathbf{x},t)}{\partial t} \ , \ \nabla \times \mathbf{E}(\mathbf{x},t) = -\frac{\partial \mathbf{B}}{\partial t} \tag{1}$$

in form a similar to the Dirac equation where the external charge and current densities are $\rho$ and **J.** For linear isotropic material media

$$\mathbf{D}(\mathbf{x},t)=\varepsilon(\mathbf{x},t)\,\mathbf{E}(\mathbf{x},t) \quad , \quad \mathbf{B}(\mathbf{x},t)=\mu(\mathbf{x},t)\,\mathbf{H}(\mathbf{x},t) \tag{2}$$

where the permittivity $\varepsilon(\mathbf{x},t)=\varepsilon_0\,\varepsilon_r(\mathbf{x},t)$ and the permeability $\mu(\mathbf{x},t)=\mu_0\,\mu_r(\mathbf{x},t)$. The speed of light in a vacuum $c=(\mu_0\,\varepsilon_0)^{-1/2}$. To rewrite the Maxwell equations into matrix form, it is convenient to introduce the two Riemann-Silberstein vectors [4, 7, 24]

$$\mathbf{F}^{\pm}=\frac{1}{\sqrt{2}}\left[\sqrt{\varepsilon}\,\mathbf{E}\pm i\frac{\mathbf{B}}{\sqrt{\mu}}\right] \tag{3}$$

In inhomogeneous media, the electromagnetic wave is a mix of the two wave polarizations and thus requires both Riemann-Silberstein vectors $\mathbf{F}^{\pm}$. In homogeneous media there is no mixing of the different wave polarizations and so only one Riemann-Silberstein vector is needed.

Following Khan [24], we introduce

$$v(\mathbf{x},t)=\frac{1}{\sqrt{\varepsilon\,\mu}}\ ,\ \ h(\mathbf{x},t)=\sqrt{\frac{\mu}{\varepsilon}} \tag{4}$$

so that the Maxwell equations in terms of the two Riemann-Silberstein vectors becomes [24]

$$i\frac{\partial\mathbf{F}^{\pm}}{\partial t}=\pm v\nabla\times\mathbf{F}^{\pm}\pm\frac{1}{2}\nabla v\times\mathbf{F}^{\pm}\pm\frac{v}{2h}\nabla h\times\mathbf{F}^{\mp}+\frac{i}{2}\left(\frac{\partial\ln v}{\partial t}\mathbf{F}^{\pm}+\frac{\partial\ln h}{\partial t}\mathbf{F}^{\mp}\right)-i\sqrt{\frac{vh}{2}}\,\mathbf{J}$$

$$\nabla\bullet\mathbf{F}^{\pm}=\frac{1}{2v}\nabla v\bullet\mathbf{F}^{\pm}+\frac{1}{2h}\nabla h\bullet\mathbf{F}^{\mp}+\sqrt{\frac{vh}{2}}\,\rho\,. \tag{5}$$

The coupling between the polarizations occur through either spatial or temporal time variations of $h(\mathbf{x},t)$. The matrix representation of Eq. (5) now takes the form [24] of 8- spinor components

$$\frac{\partial}{\partial t}\begin{pmatrix}\Psi^{+}\\ \Psi^{-}\end{pmatrix}-\frac{1}{2}\frac{\partial\ln v}{\partial t}\begin{pmatrix}\Psi^{+}\\ \Psi^{-}\end{pmatrix}+\frac{iM_z\alpha_y}{2}\frac{\partial\ln h}{\partial t}\begin{pmatrix}\Psi^{-}\\ \Psi^{-}\end{pmatrix}$$

$$=-v\begin{pmatrix}\mathbf{M}\bullet\nabla+\vec{\Sigma}\bullet\dfrac{\nabla v}{2v} & -iM_z\vec{\Sigma}\bullet\dfrac{\nabla h}{h}\alpha_y\\ -iM_z\vec{\Sigma}^{*}\bullet\dfrac{\nabla h}{h}\alpha_y & \mathbf{M}^{*}\bullet\nabla+\vec{\Sigma}^{*}\bullet\dfrac{\nabla v}{2v}\end{pmatrix}\begin{pmatrix}\Psi^{+}\\ \Psi^{-}\end{pmatrix}-\begin{pmatrix}W^{+}\\ W^{-}\end{pmatrix} \tag{6}$$

where the Cartesian Riemann-Silberstein components and source matrices are defined by

$$\Psi^{\pm}=\begin{pmatrix}-F_x^{\pm}\pm i\,F_y^{\pm}\\ F_z^{\pm}\\ F_z^{\pm}\\ F_x^{\pm}\pm i\,F_y^{\pm}\end{pmatrix}\quad,\quad W^{\pm}=\frac{1}{\sqrt{2\varepsilon}}\begin{pmatrix}-J_x\pm i\,J_y\\ J_z-v\,\rho\\ J_z+v\,\rho\\ J_x\pm i\,J_y\end{pmatrix} \tag{7}$$

On using the spin ½ Pauli spin matrices

$$\sigma_x=\begin{pmatrix}0 & 1\\ 1 & 0\end{pmatrix}\quad,\quad\sigma_y=\begin{pmatrix}0 & -i\\ i & 0\end{pmatrix}\ ,\ \ \sigma_z=\begin{pmatrix}1 & 0\\ 0 & -1\end{pmatrix} \tag{8}$$

the 4x4 matrices **M** in Eq. (6) are just the tensor product of the Pauli matrices with the 2x2 identity matrix $I_2 : \mathbf{M} = \vec{\sigma} \otimes I_2$, with $M_z = \sigma_z \otimes I_2$ . Finally

$$\vec{\alpha} = \begin{pmatrix} 0 & \vec{\sigma} \\ \vec{\sigma} & 0 \end{pmatrix} \text{and} \quad \vec{\Sigma} = \begin{pmatrix} \vec{\sigma} & 0 \\ 0 & \vec{\sigma} \end{pmatrix} \quad . \tag{9}$$

For homogeneous media, $\nabla v = 0 = \nabla h = \partial v / \partial t = \partial h / \partial t$ , so that Eq. (6) decouples to

$$\frac{\partial \Psi^+}{\partial t} = -v \mathbf{M} \bullet \nabla \Psi^+ - W^+ \tag{10}$$

The sum of the 1st and 4th rows of Eq. (10) determines the time evolution of $F_y$, i.e., of the y-components of **E** and **B,** while the difference of the 1st and 4th rows yields time evolution of the x-component of **E** and **B.** The sum of the 2nd and 3rd rows of Eq. (10) will yield the time evolution of the z-component of **E** and **B.** Thus we have determined the time evolution parts of Maxwell's equations. The divergence equations of the Maxwell equations will come from taking the difference of the 2nd and 3rd rows of Eq. (10).

Alternatively, Jetstadt et. al. [7] restrict themselves to the 3-spinor components

$$\Phi^{\pm} = \begin{pmatrix} -F_x^{\pm} + i F_y^{\pm} \\ F_z^{\pm} \\ F_x^{\pm} + i F_y^{\pm} \end{pmatrix} \tag{11}$$

In this representation it is clear that one will only recover the time-dependent parts of Maxwell's equations and not the divergence equations $\nabla \bullet \mathbf{D} = \rho$ , $\nabla \bullet \mathbf{B} = 0$ . These will have to be imposed as constraints.

## II. Unitary Quantum Lattice Algorithm

### A. Dirac Equation

What drew the attention of researchers from as early as 1931 was the similarity between the Riemann-Silberstein vector representation of Maxwell equations and the Dirac equation. One form of the Dirac equation for a free particle of mass *m* is the 4-spinor evolution of $\psi$

$$\frac{\partial \psi}{\partial t} = c \sum_{j=1}^{3} a \otimes \sigma_j \frac{\partial \psi}{\partial x_j} + i b \otimes I_2 m \psi \tag{12}$$

where $a$ and $b$ are any Pauli spin matrices, but $a \neq b$ . In particular [5,6] for the choice $a = \sigma_x$ , $m = 0$ , and suitable normalization, Eq. (12) for a massless particle reduces to

$$\frac{\partial}{\partial t}\begin{pmatrix} \psi_0 \\ \psi_1 \\ \psi_2 \\ \psi_3 \end{pmatrix} = \frac{\partial}{\partial x}\begin{pmatrix} \psi_3 \\ \psi_2 \\ \psi_1 \\ \psi_0 \end{pmatrix} + i \frac{\partial}{\partial y}\begin{pmatrix} -\psi_3 \\ \psi_2 \\ -\psi_1 \\ \psi_0 \end{pmatrix} + \frac{\partial}{\partial z}\begin{pmatrix} \psi_2 \\ -\psi_3 \\ \psi_0 \\ -\psi_1 \end{pmatrix} \quad . \tag{13}$$

### B. Maxwell's Equations in Homogeneous Media for Propagation in 2D

In Eq. (7), one needs only introduce the 4-spinor components $\{q_0, q_1, q_2, q_3\}$:

$$\mathbf{\Psi}^{+}=\begin{pmatrix}-F_x^{+}+iF_y^{+}\\ F_z^{+}\\ F_z^{+}\\ F_x^{+}+iF_y^{+}\end{pmatrix}\equiv\begin{pmatrix}q_0\\ q_1\\ q_2\\ q_3\end{pmatrix}, \tag{14}$$

Eq. (10) for homogeneous media (and with no external sources) reduces to (on setting c = 1)

$$\frac{\partial}{\partial t}\begin{pmatrix}q_0\\ q_1\\ q_2\\ q_3\end{pmatrix}=-\frac{\partial}{\partial x}\begin{pmatrix}q_2\\ q_3\\ q_0\\ q_1\end{pmatrix}+i\frac{\partial}{\partial y}\begin{pmatrix}q_2\\ q_3\\ -q_0\\ -q_1\end{pmatrix}-\frac{\partial}{\partial z}\begin{pmatrix}q_0\\ q_1\\ -q_2\\ -q_3\end{pmatrix} \tag{15}$$

The similarity between the Dirac equation for a massless particle, Eq. (13), and Maxwell equations in a homogeneous medium, Eq. (15), is clear.

A QLA for the Maxwell equations, Eq. (15), can now be readily determined, building on the Dirac-QLA of Yepez [5,6]. Here we will concentrate on determining such an algorithm for 1D and 2D Maxwell equations – while the 3D version will be addressed in the future. We introduce the 4-spinor components, Eq. (14), and consider the unitary collision operators

$$C_X=\begin{pmatrix}\cos\theta & 0 & \sin\theta & 0\\ 0 & \cos\theta & 0 & \sin\theta\\ -\sin\theta & 0 & \cos\theta & 0\\ 0 & -\sin\theta & 0 & \cos\theta\end{pmatrix},\quad C_Y=\begin{pmatrix}\cos\theta & 0 & i\sin\theta & 0\\ 0 & \cos\theta & 0 & i\sin\theta\\ i\sin\theta & 0 & \cos\theta & 0\\ 0 & i\sin\theta & 0 & \cos\theta\end{pmatrix} \tag{16}$$

and the unitary streaming operators $S_{\pm X}^{01}, S_{\pm X}^{23}$ which shift the appropriate amplitudes $\left\{q_j\left(\mathbf{x},t\right), j=0,3\right\}$ along the lattice in the x-direction by $\pm 1$ lattice units:

$$S_{\pm X}^{01}\begin{pmatrix}q_0(x,y,t)\\ q_1(x,y,t)\\ q_2(x,y,t)\\ q_3(x,y,t)\end{pmatrix}=\begin{pmatrix}q_0(x\pm 1,y,t)\\ q_1(x\pm 1,y,t)\\ q_2(x,y,t)\\ q_3(x,y,t)\end{pmatrix},\quad S_{\pm X}^{23}\begin{pmatrix}q_0(x,y,t)\\ q_1(x,y,t)\\ q_2(x,y,t)\\ q_3(x,y,t)\end{pmatrix}=\begin{pmatrix}q_0(x,y,t)\\ q_1(x,y,t)\\ q_2(x\pm 1,y,t)\\ q_3(x\pm 1,y,t)\end{pmatrix}. \tag{17}$$

There are similar expressions for the unitary streaming operators in the y-direction: $S_{\pm Y}^{01}, S_{\pm Y}^{23}$. For the x-direction, one now considers the following interleaved sequence of unitary operators:

$$\begin{aligned}U_X &= S_{-X}^{01}C_X S_{+X}^{01}C_X^{\dagger}.S_{+X}^{23}C_X S_{-X}^{23}C_X^{\dagger}\\ U_X^{adj} &= S_{+X}^{01}C_X^{\dagger}S_{-X}^{01}C_X.S_{-X}^{23}C_X^{\dagger}S_{+X}^{23}C_X\end{aligned} \tag{18}$$

while for the y-direction we consider a slightly different sequence of interleaved operators

$$\begin{aligned}U_Y &= S_{-Y}^{23}C_Y S_{+Y}^{23}C_Y^{\dagger}.S_{+Y}^{01}C_Y S_{-Y}^{01}C_Y^{\dagger}\\ U_Y^{adj} &= S_{+Y}^{23}C_Y^{\dagger}S_{-Y}^{23}C_Y.S_{+Y}^{01}C_Y^{\dagger}S_{-Y}^{01}C_Y\end{aligned}. \tag{19}$$

The time advancement of the 4-spinor components

$$\begin{pmatrix} q_0 \\ q_1 \\ q_2 \\ q_3 \end{pmatrix}_{t+\delta t} = U_Y^{adj} U_Y U_X^{adj} U_X \begin{pmatrix} q_0 \\ q_1 \\ q_2 \\ q_3 \end{pmatrix}_t \tag{20}$$

yields the following evolution of the 4-spinor amplitudes (on using Mathematica) under diffusion ordering, on scaling the time advancement $\delta t = \varepsilon^2$ and lattice spacing $\delta x = \delta y = \varepsilon$ ,

$$\frac{\partial}{\partial t}\begin{pmatrix} q_0 \\ q_1 \\ q_2 \\ q_3 \end{pmatrix} = -\frac{\partial}{\partial x}\begin{pmatrix} q_2 \\ q_3 \\ q_0 \\ q_1 \end{pmatrix} + i\frac{\partial}{\partial y}\begin{pmatrix} q_2 \\ q_3 \\ -q_0 \\ -q_1 \end{pmatrix} + O(\varepsilon^2) \tag{21}$$

provided the collision angle in Eq. (16) is $\theta = \varepsilon / 4$. Equation (21) is just the Maxwell equations for electromagnetic fields with 2D spatial dependence, and Eq. (20) is its QLA representation.

**C. Some QLA Simulations for the 2D Maxwell Equations in a Vacuum**

First, we shall consider a Gaussian pulse propagating in the y-direction with initial condition

$$E_z(x,y,t=0) = E_0 \exp\left[-\frac{(y-y_0)^2}{\sigma^2}\right] \cos\left[k_y (y-y_0)\right]$$
$$B_x(x,y,t=0) = E_z(x,y,t=0) \tag{22}$$

with all the other field components zero: $E_x = 0 = E_y = B_y = B_z$ at *t = 0*. The QLA algorithm, Eq. (20), is solved on a 5000 x 5000 grid, with the small parameter $\varepsilon = 0.1$ in the collision angle $\theta = \varepsilon / 4$ . For parameters $E_0 = 0.01$, $\sigma^2 = 9000$ , $k_y = 0.08$, the initial Gaussian pulse for $E_z$ is

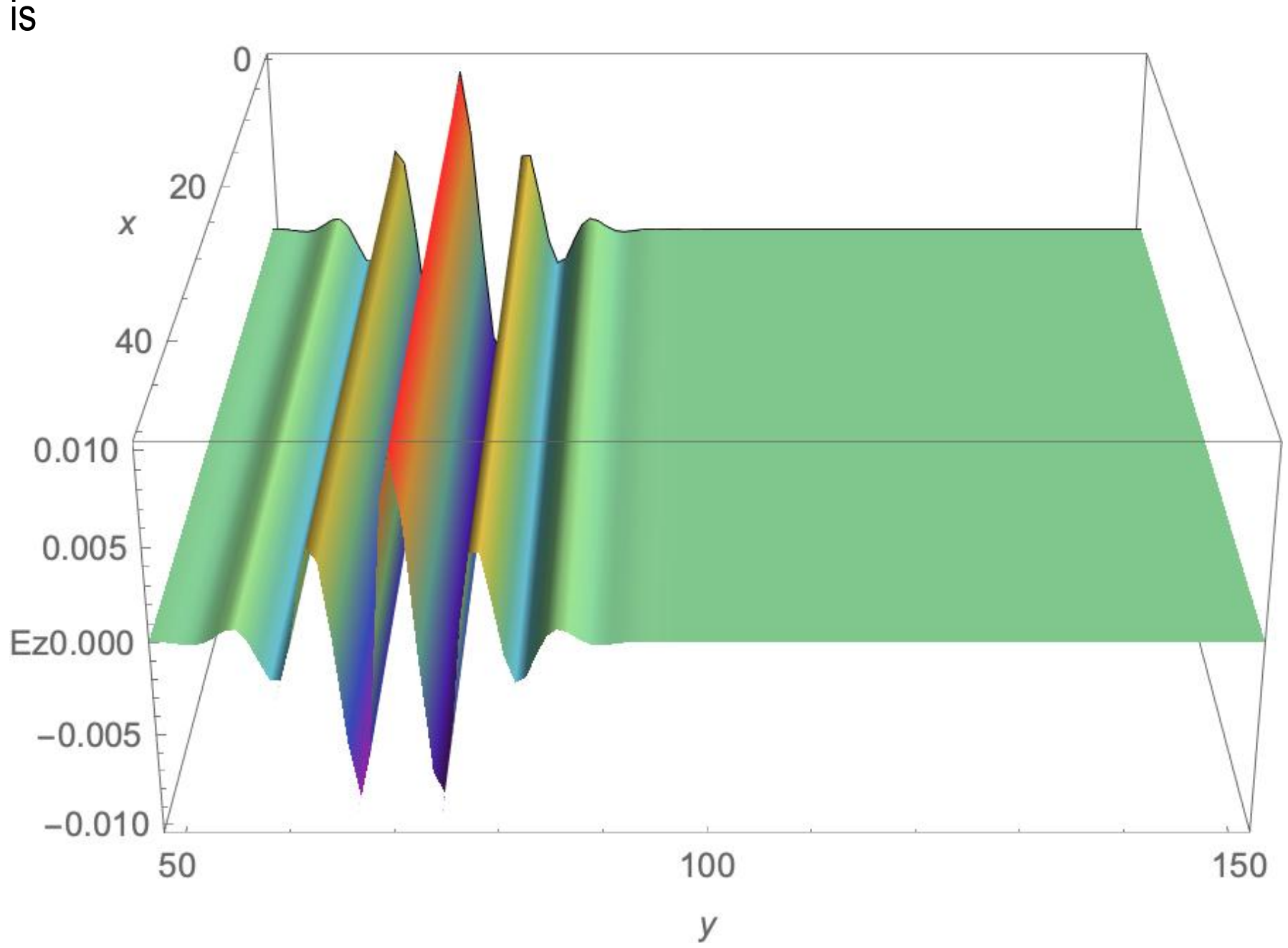

*Fig. 1 The initial Gaussian pulse for the electric field component $E_z(t=0)$, plotted at every tenth point in the x- and y- directions. (i.e., the actual simulation data grid is 500 < y' < 1500, 0 < x' < 500).*

shown in Fig. 1. with $E_z(x,y,t)$ being determined by the symmetrized form

$$E_z = 0.5\ \mathrm{Re}\left[q_1 + q_2\right] \tag{23}$$

since, from Eq. (14) , $q_1 = q_2 = F_z^+ = E_z + i\,B_z$. After 1000 time steps, the wave packet has propagated along the y-axis undistorted, (c = 1) Fig. 2:

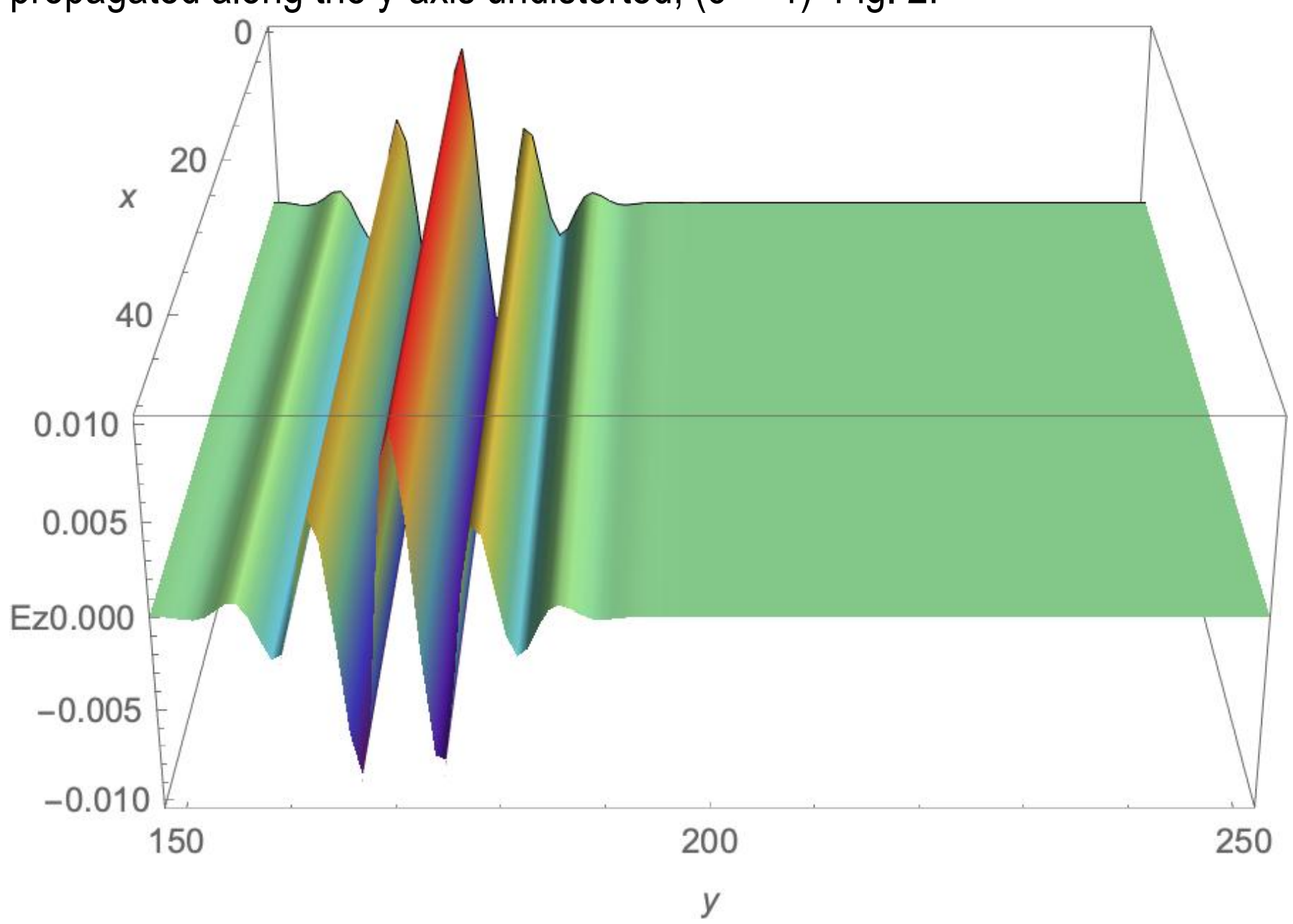

*Fig. 2 Gaussian pulse $E_z$ at time t = 1k.*

After t = 30k time steps, under periodic boundary conditions, there is no discernable distortion in the pulse shape , Fig. 3

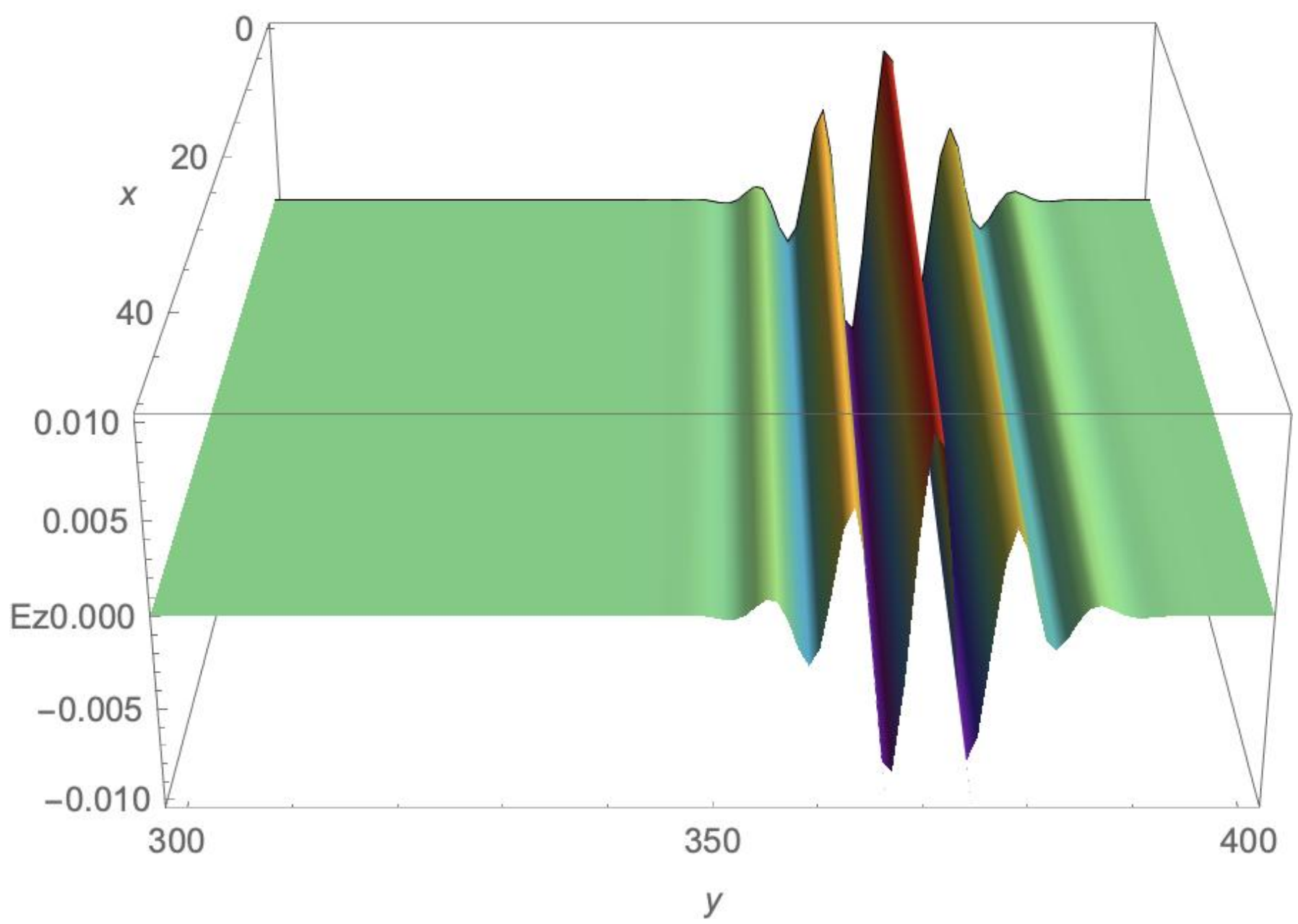

*Fig. 3 Guassian pulse $E_z$ at t = 30k.*

The noise level is 7 orders of magnitude lower than the pulse amplitude, as seen in Fig. 4 and 5

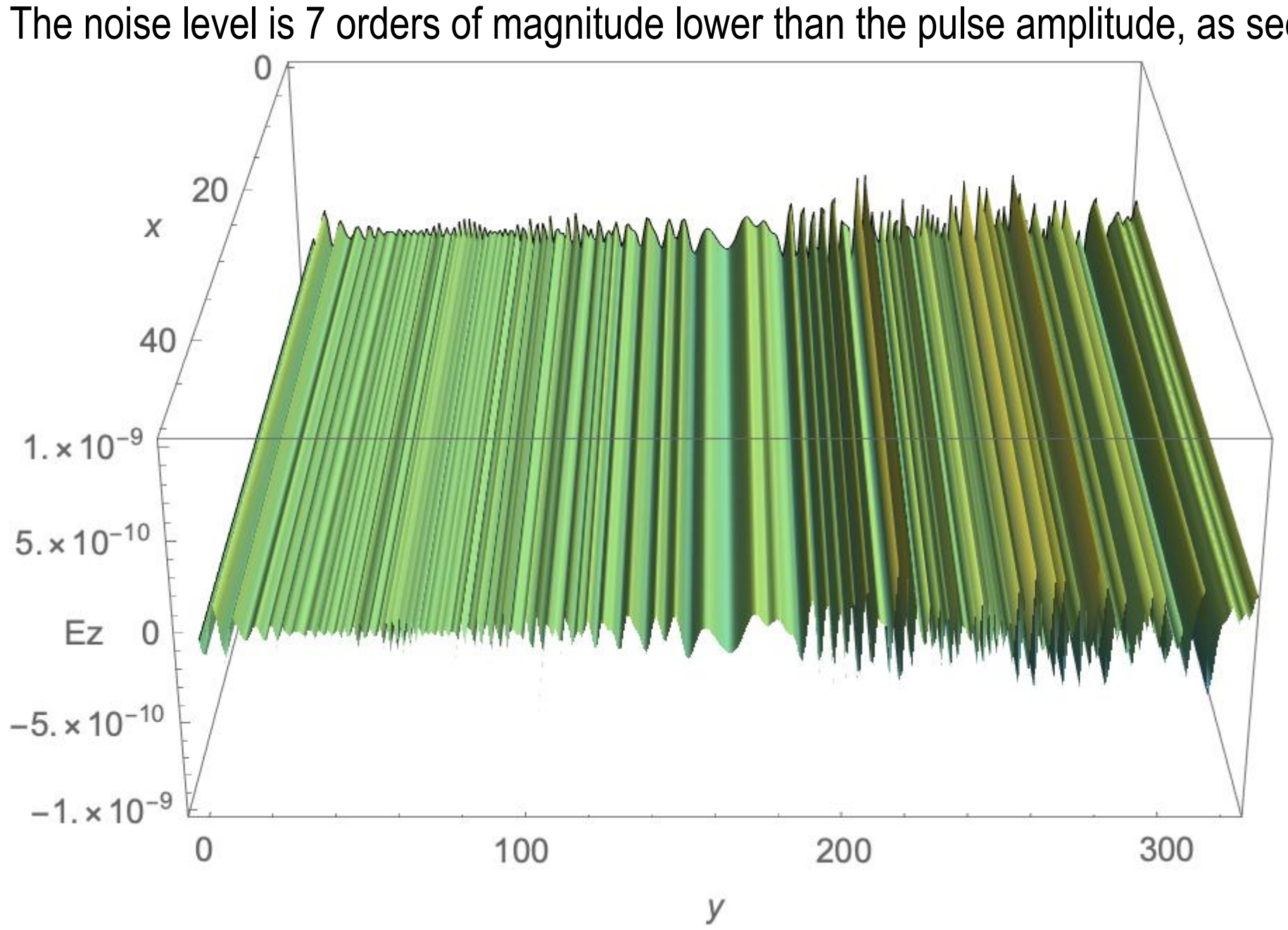

*Fig. 4 The noise in the electric field $E_z$ in the spatial region before the Gaussian pulse ( Fig. 3) at time t = 30k. Note that the noise is 7 orders of magnitude below the peak in the Gaussian pulse.*

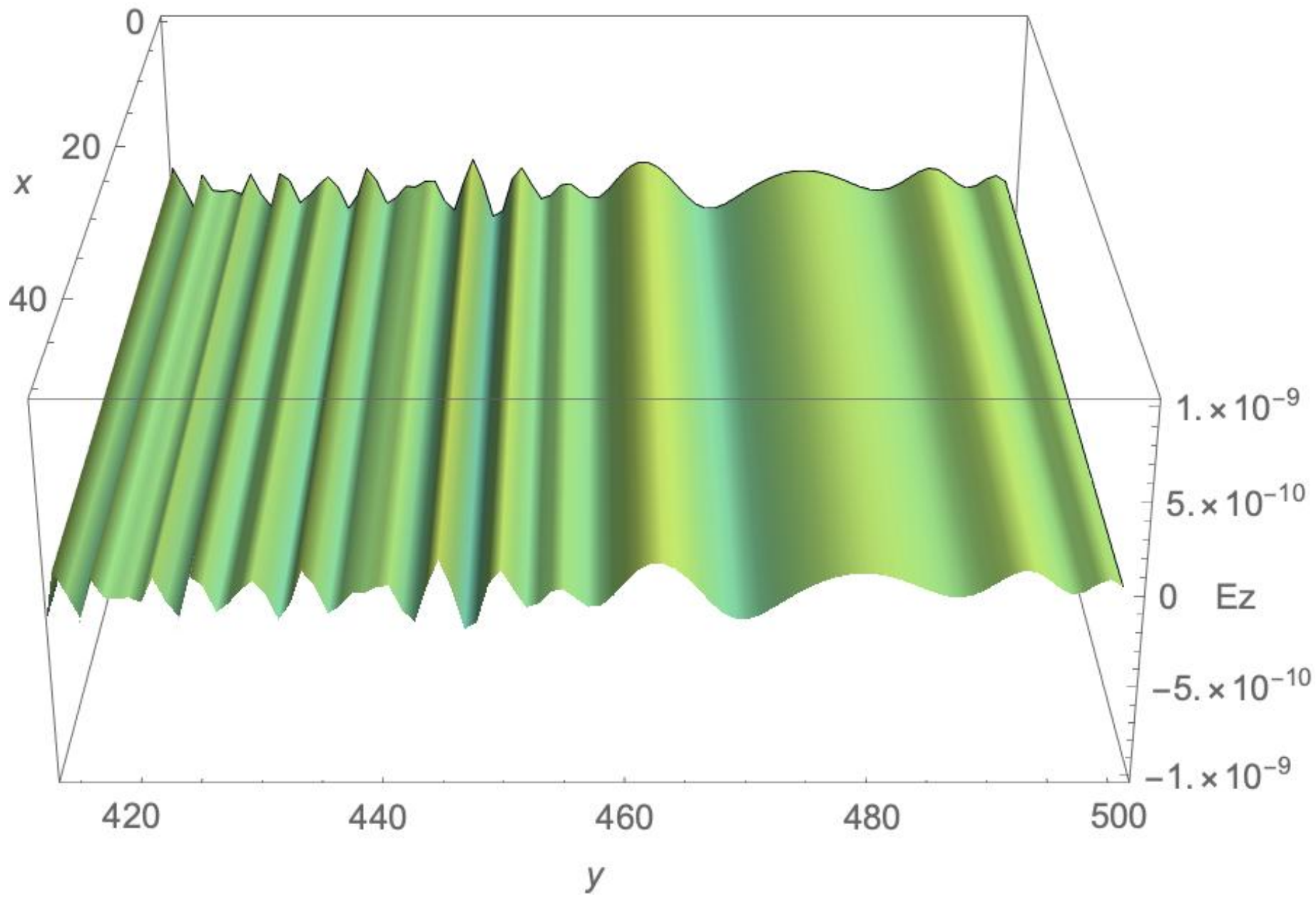

*Fig. 5 The noise level in the electric field $E_z$ in the spatial region after the Gaussian pulse ( Fig. 3) at time t = 30k. The noise is 7 orders of magnitude below the peak in the Gaussian pulse*

Because the evolution equations for the spinor amplitudes $q_1$ and $q_2$ are different in Eq. (21), it is interesting to the plot the difference $\mathrm{Re}\left[q_1 - q_2\right]$. This is shown in Fig. 6

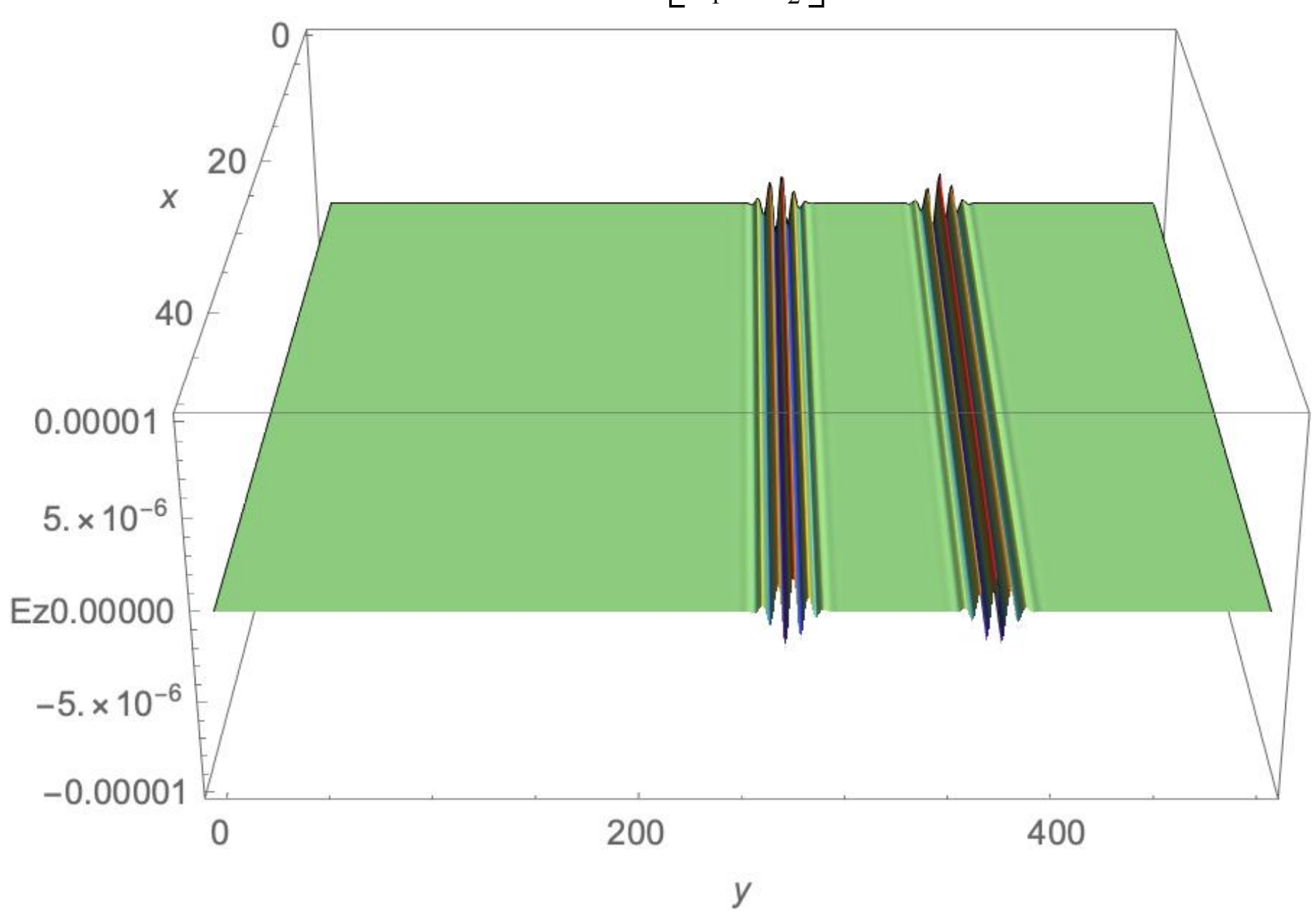

*Fig. 6 A plot of the difference in the spinor amplitudes $\mathrm{Re}[q_1 - q_2]$ at t = 30k.*

Finally we show the Gaussian pulse after 130 k iterations, Fig. 7

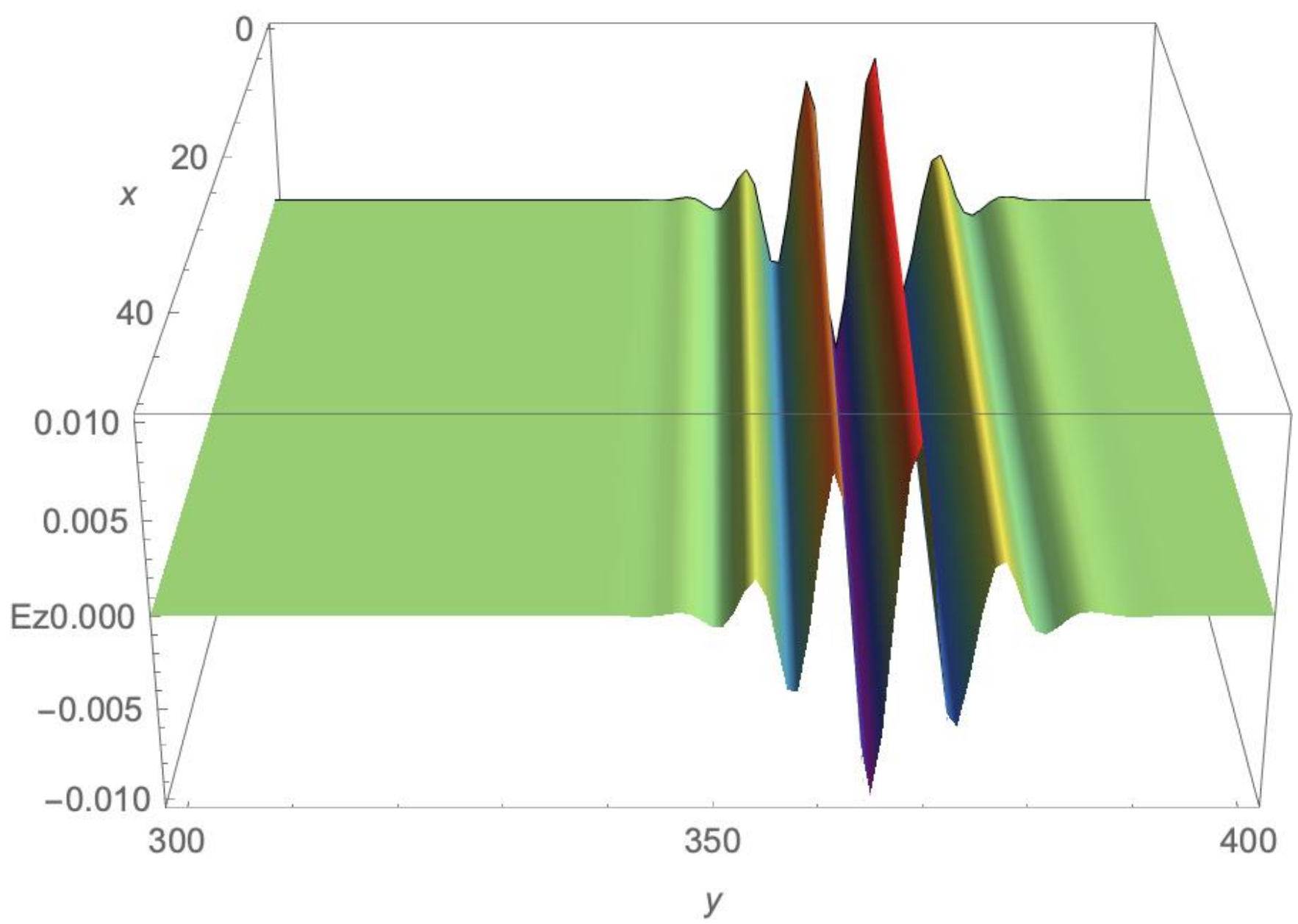

*Fig. 7 The $E_z$-component of the Gaussian pulse at t = 130 K.*

while the signal before (Fig. 8) and after (Fig. 9) the Gaussian pulse, Fig. 7, remain more than 7 orders of magnitude lower in amplitude.

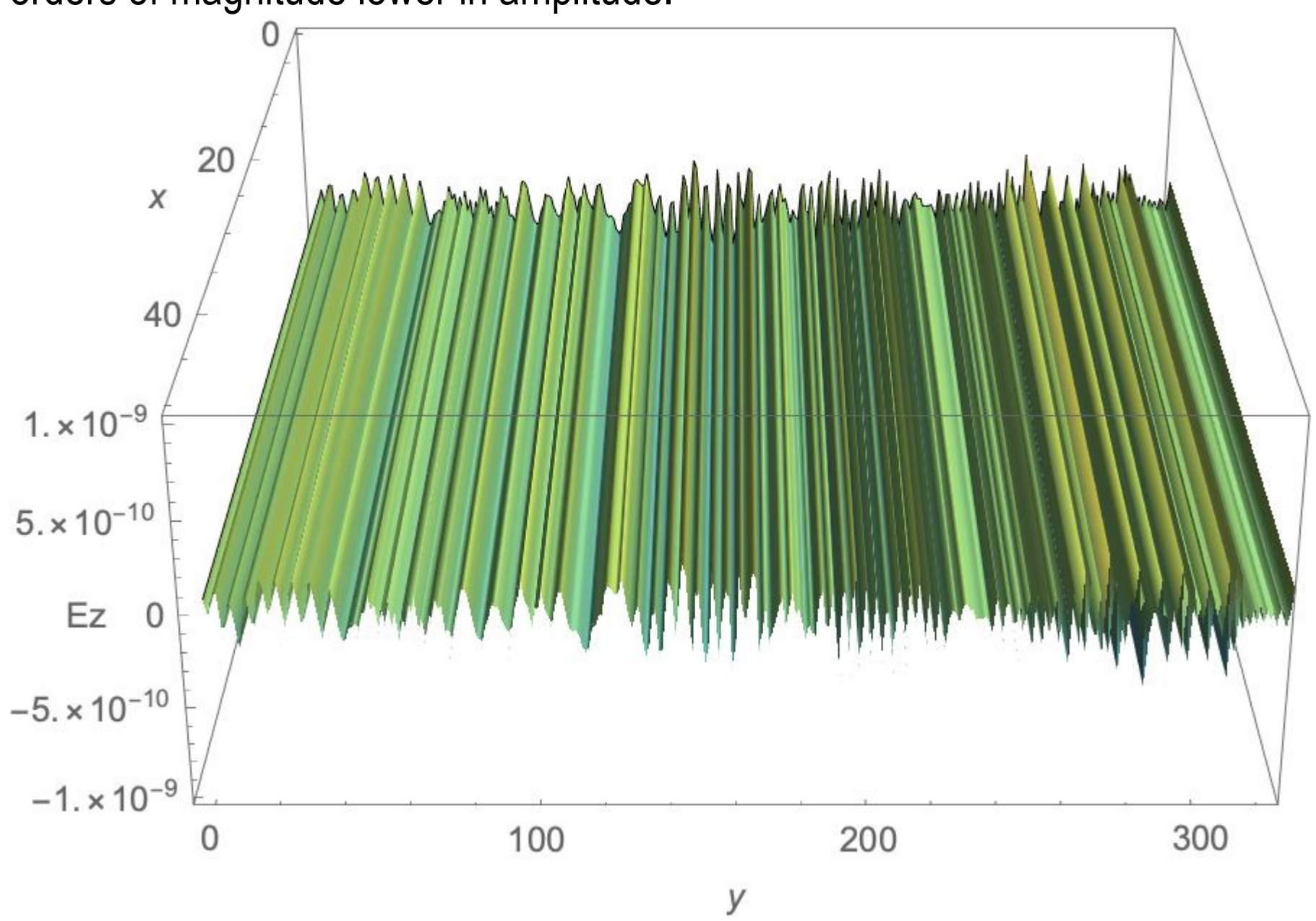

*Fig. 8 The $E_z$-component of the Gaussian pulse for spatial regions before the pulse at t = 130 K. These signal strengths remain over 7 orders of magnitude below that of the main Gaussian pulse.*

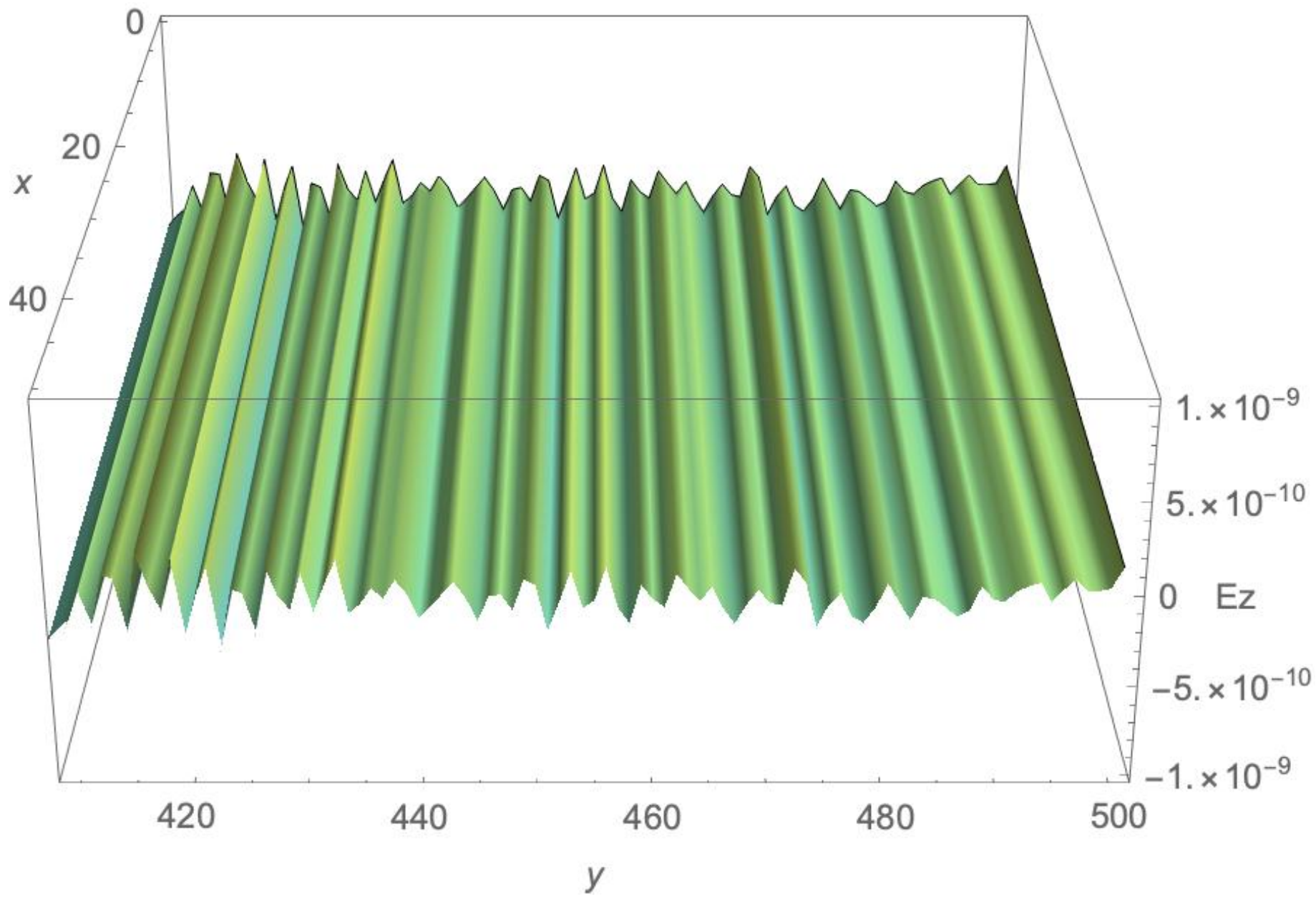

*Fig. 9 The $E_z$-component of the Gaussian pulse for spatial regions before the pulse at t = 130 K. These signal strengths remain over 7 orders of magnitude below that of the main Gaussian pulse.*

**Pulse-propagation in the x-direction**

We obtain similar results for pulse propagation in the y-direction, even though the collide-stream sequence $U_X$ is significantly different from that required in $U_Y$. This difference in these unitary sequences reflects the lack of symmetry in x-y interchange in the Dirac formulation of Maxwell's equations, Eq. (21).

## III. UNITARY QLA SIMULATIONS FOR 1D MAXWELL'S EQUATIONS IN INHOMOGENEOUS MEDIA

We now consider the case of normal incidence of an electromagnetic wave onto a dielectric boundary, permitting only a spatial dependence in y. i.e., we consider an electromagnetic wave with non-zero components $E_z$ , $B_x$ with refractive index $n(y)$. Equations (6)-(8) reduce to the following 8-spinor representation which is conveniently written in two blocks of 4-spinor components:

$$\frac{\partial}{\partial t}\begin{pmatrix} q_0 \\ q_1 \\ q_2 \\ q_3 \end{pmatrix} = \frac{1}{n(y)} i \frac{\partial}{\partial y}\begin{pmatrix} q_2 \\ q_3 \\ -q_0 \\ -q_1 \end{pmatrix} - i\frac{n'(y)}{2n^2(y)}\begin{pmatrix} q_1 - q_6 \\ -q_0 - q_7 \\ q_3 + q_4 \\ -q_2 + q_5 \end{pmatrix}$$

$$\frac{\partial}{\partial t}\begin{pmatrix} q_4 \\ q_5 \\ q_6 \\ q_7 \end{pmatrix} = \frac{1}{n(y)} i \frac{\partial}{\partial y}\begin{pmatrix} -q_6 \\ -q_7 \\ q_4 \\ q_5 \end{pmatrix} - i\frac{n'(y)}{2n^2(y)}\begin{pmatrix} -q_5 - q_2 \\ q_4 - q_3 \\ -q_7 + q_0 \\ q_6 + q_1 \end{pmatrix} \tag{24}$$

where $n'(y) = dn/dy$ , and

$$\begin{pmatrix} q_0 \\ q_1 \\ q_2 \\ q_3 \\ q_4 \\ q_5 \\ q_6 \\ q_7 \end{pmatrix} = \begin{pmatrix} -F_x^+ + i F_y^+ \\ F_z^+ \\ F_z^+ \\ F_x^+ + i F_y^+ \\ -F_x^- - i F_y^- \\ F_z^- \\ F_z^- \\ F_x^- - i F_y^- \end{pmatrix} \quad , \quad with \ \mathbf{F}^{\pm} = \frac{1}{\sqrt{2}}\left[ \sqrt{\varepsilon}\,\mathbf{E} \pm i \frac{\mathbf{B}}{\sqrt{\mu}} \right]. \tag{25}$$

The two Riemann-Silberstein vectors for the two different polarizations, $\mathbf{F}^+ \text{ and } \mathbf{F}^-$, are coupled by the spatial gradient in the refractive index $n(y) = \sqrt{\mu_0\, \varepsilon(y)}$. For simplicity, we shall consider normal incidence from a region of constant dielectric $n_0$ to a region of higher constant dielectric $n_1$:

$$n(y) = \begin{cases} n_0 & , \ y < L_1 \\ n_1 > n_0 & , \ y > L_1 \end{cases} \tag{26}$$

The QLA to reproduce the 1D Maxwell equations, Eq. (24), to second order accuracy has the following unitary collision operator

$$C_Y(\theta) = \begin{pmatrix} \cos\theta & 0 & i\sin\theta & 0 & 0 & 0 & 0 & 0 \\ 0 & \cos\theta & 0 & i\sin\theta & 0 & 0 & 0 & 0 \\ i\sin\theta & 0 & \cos\theta & 0 & 0 & 0 & 0 & 0 \\ 0 & i\sin\theta & 0 & \cos\theta & 0 & 0 & 0 & 0 \\ 0 & 0 & 0 & 0 & \cos\theta & 0 & -i\sin\theta & 0 \\ 0 & 0 & 0 & 0 & 0 & \cos\theta & 0 & -i\sin\theta \\ 0 & 0 & 0 & 0 & -i\sin\theta & 0 & \cos\theta & 0 \\ 0 & 0 & 0 & 0 & 0 & -i\sin\theta & 0 & \cos\theta \end{pmatrix} \tag{27}$$

which interleaved with the unitary streaming operator will recover the 1st term on the right-hand side of Eq. (24) provided

$$\theta = \frac{\varepsilon}{4n(y)} \quad . \tag{28}$$

To the recover the inhomogeneous dielectric factor $n'(y)$ in Eq. (24) one introduces the Hermitian operators

$$V_{11} = \begin{pmatrix} \cos\alpha & \sin\alpha & 0 & 0 & 0 & 0 & 0 & 0 \\ -\sin\alpha & \cos\alpha & 0 & 0 & 0 & 0 & 0 & 0 \\ 0 & 0 & \cos\alpha & \sin\alpha & 0 & 0 & 0 & 0 \\ 0 & 0 & -\sin\alpha & \cos\alpha & 0 & 0 & 0 & 0 \\ 0 & 0 & 0 & 0 & \cos\alpha & -\sin\alpha & 0 & 0 \\ 0 & 0 & 0 & 0 & \sin\alpha & \cos\alpha & 0 & 0 \\ 0 & 0 & 0 & 0 & 0 & 0 & \cos\alpha & -\sin\alpha \\ 0 & 0 & 0 & 0 & 0 & 0 & \sin\alpha & \cos\alpha \end{pmatrix}$$

$$V_{22} = \begin{pmatrix} \cos\alpha & 0 & 0 & 0 & 0 & 0 & -\sin\alpha & 0 \\ 0 & \cos\alpha & 0 & 0 & 0 & 0 & 0 & -\sin\alpha \\ 0 & 0 & \cos\alpha & 0 & \sin\alpha & 0 & 0 & 0 \\ 0 & 0 & 0 & \cos\alpha & 0 & \sin\alpha & 0 & 0 \\ 0 & 0 & -\sin\alpha & 0 & \cos\alpha & 0 & 0 & 0 \\ 0 & 0 & 0 & -\sin\alpha & 0 & \cos\alpha & 0 & 0 \\ \sin\alpha & 0 & 0 & 0 & 0 & 0 & \cos\alpha & 0 \\ 0 & \sin\alpha & 0 & 0 & 0 & 0 & 0 & \cos\alpha \end{pmatrix} \tag{29}$$

with the rotation angle

$$\alpha = -i\varepsilon^2 \frac{n'(y)}{2n^2(y)} . \tag{30}$$

Now each of these Hermitian matrices can be decomposed into a sum of two unitary matrices : e.g., on normalizing $V_{11}$ so that $\|V_{11}\| \le 1$ then one can rewrite

$$V_{11} = \frac{1}{2}\left(U_{11}^{(1)} + U_{11}^{(2)}\right) \tag{31}$$

where $U_{11}^{(1)}$ and $U_{11}^{(2)}$ are now unitary

$$U_{11}^{(1)} = V_{11} + i\sqrt{I - V_{11}^2} \quad , \quad U_{11}^{(2)} = V_{11} - i\sqrt{I - V_{11}^2} \tag{32}$$

Childs & Wiebe [29] have shown that one can encode linear combinations of unitary operators on quantum computers and that in some cases these algorithms will outperform the usual product of unitary operators algorithms.

The unitary interleaved sequence of collide-stream-potential operators

$$\begin{gathered} U_{YY} = S_{-Y}^{23,67} C_Y(\theta) S_{+Y}^{23,67} C_Y^\dagger(\theta) . S_{+Y}^{01,45} C_Y(\theta) S_{-X}^{01,45} C_Y^\dagger(\theta) \\ U_{YY}^{adj} = S_{+Y}^{23,67} C_Y^\dagger(\theta) S_{-Y}^{23,67} C_Y(\theta) . S_{+Y}^{01,45} C_Y^\dagger(\theta) S_{-Y}^{01,45} C_Y(\theta) \end{gathered} , \quad \text{with } \theta = \frac{\varepsilon}{4n(y)} \tag{34}$$

are thus augmented with the Hermitian operators that can be decomposed into a sum of two unitary matrices which can still be encoded onto a quantum computer

$$\vec{\mathbf{q}}(t+\delta t) = V_{22} V_{11} U_{YY}^{adj} U_{YY} \ \vec{\mathbf{q}}(t) \tag{35}$$

where $\vec{\mathbf{q}}$ is the 8-spinor, Eq. (25).

We have performed some 1D simulations of electromagnetic wave propagation from a region of refractive index $n_0$ into a region with refractive index $n_1$. The refractive index profile is modeled by the hyperbolic tangent function-profile

$$n(y) = \frac{n_0 + n_1}{2} - \frac{n_0 - n_1}{2} \tanh\left(\beta\left[y - L\right]\right) \tag{36}$$

where $\beta$ controls the thickness of the boundary region between the two media. Some care needs to be taken with the perturbation parameter $\varepsilon$, as the collide-stream unitary operators have $\theta = O(\varepsilon)$ while the unitary operators controlling the media refractive interface have $\alpha = O(\varepsilon^2)$. For the simulations reported here, the boundary region between the two media is centered at $L_m = 16000$ (lattice units) with the end of the grid at $L_{end} = 2L_m$. Periodic boundary conditions are enforced by adding a small buffer region after $L_{end}$ so that the refractive index is periodic as shown in Fig. 10(a)

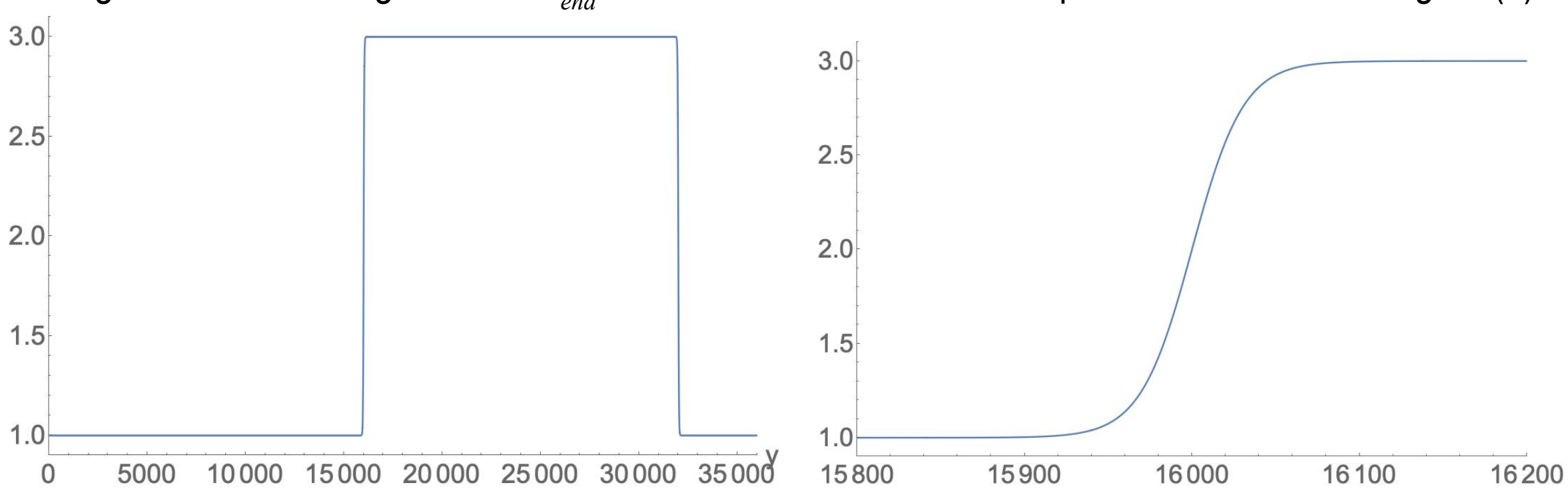

*Fig. 10 (a) refractive index profile (b) blow-up of the boundary region between media*

First, consider a simple pulse propagating from the region of refractive index $n_0 = 1$ towards the region with $n_1 = 3$. $\varepsilon = 0.3$. The initial electric and magnetic field profiles are chosen to be solutions of the Maxwell equations with $B_x(y,0) = n(y) E_z(y,0)$. Hence, when propagating in the vacuum region the $E_z$ and $B_x$ profiles overlap, Fig. 11(a) and (b), where t = 20,000 time iteration:

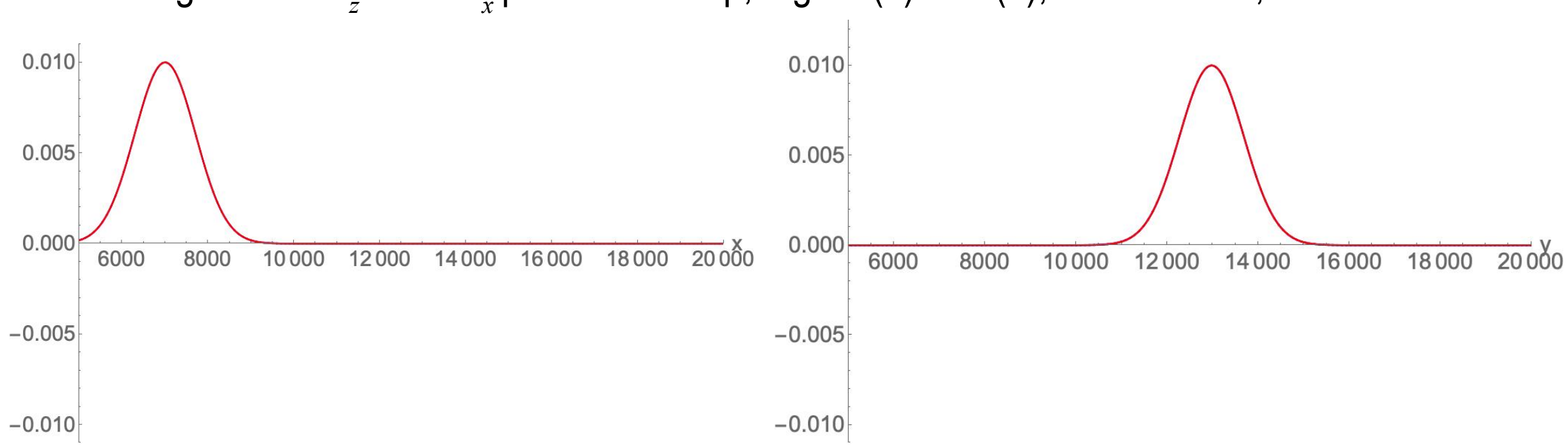

*(a) t = 0* *(b) t = 20k*

*Fig. 11 The propagation of an electromagnetic pulse from vacuum into a dielectric region with $n_1 = 3$. Interface at y = 16000. Initially the field components overlap : $E_z(blue)$, $B_x(red)$. The unitary QLA reproduces the Maxwell equations as the profiles propagate undistorted in the vacuum, as seen after 20,000 time iterations (t = 20k), (b).*

In Fig. 12 (t = 28k) the pulse has reached the interface with the separation of the electromagnetic $E_z(blue)$, $B_x(red)$ fields is clearly visible. Moreover, at t = 32k, we notice that the $E_z(blue)$

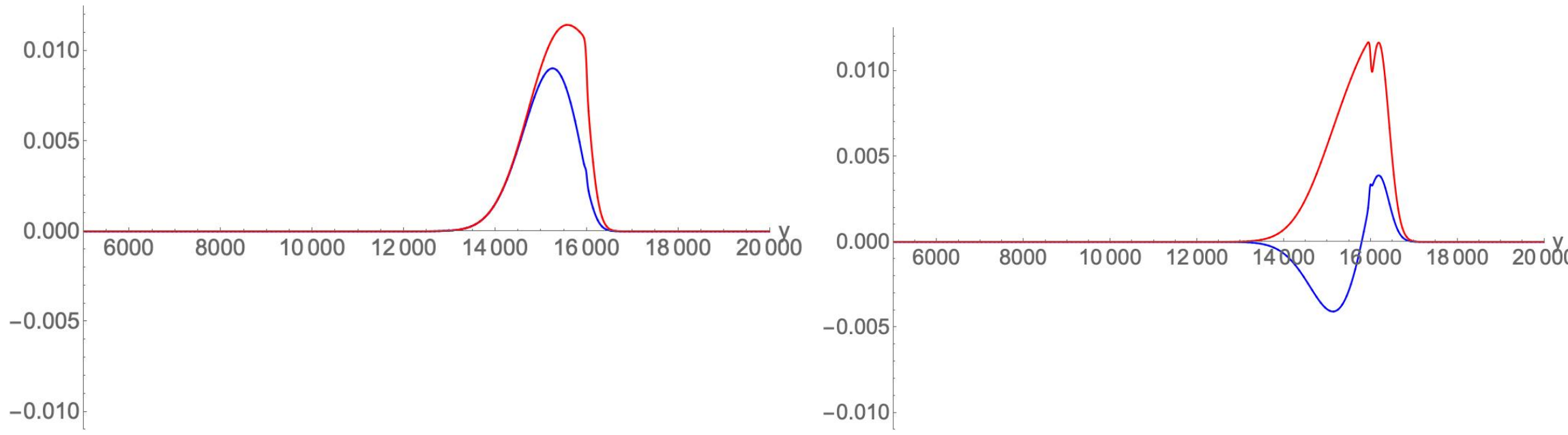

(a) t = 28K (b) t = 32K

*Fig. 12 The effect of the boundary region on the pulse as it is straddles the two media. (a) t = 28k : a separation* starts to occur between the $E_z(blue)$, $B_x(red)$ fields, with (b) showing an inversion in the $E_z(blue)$ at t = 32k.

field undergoes a $\pi$ -phase change for y < 16000. This is in accordance with standard electromagnetic theory for a plane wave incident on a dielectric interface: the ratio of the reflected to incident electric field $E_{refl} / E_{inc} = (n_0 - n_1)/(n_0 + n_1)$ with $n_0 = 1$, $n_1 = 3$

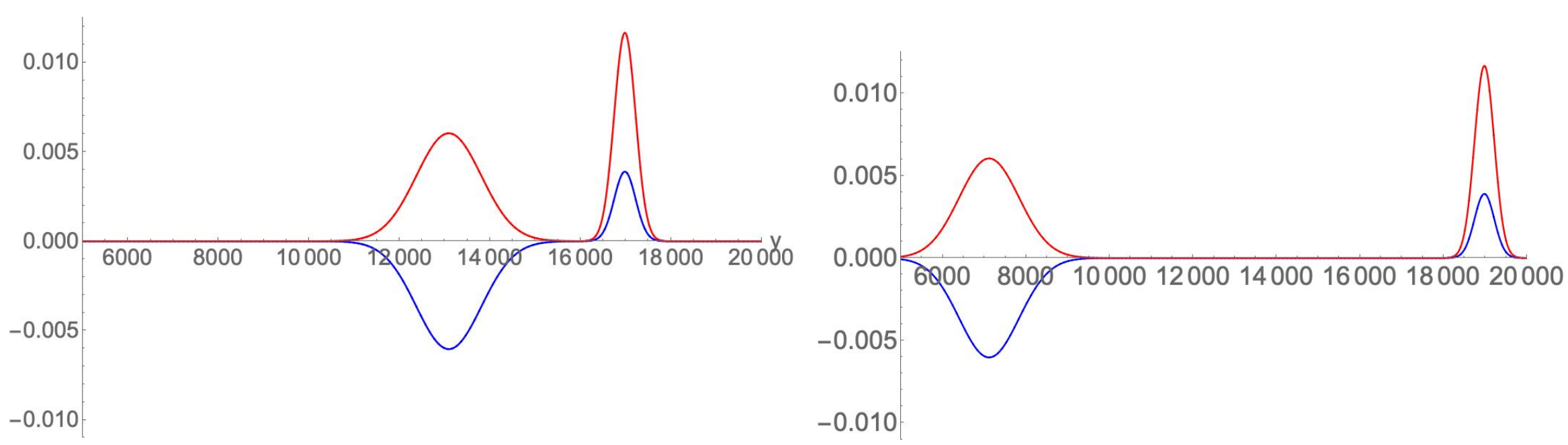

*(a) t = 40k* *(b) t = 60k*

*Fig. 13 The reflected and transmitted fields. The reflected* $E_z(blue)$ *fields suffers a* $\pi$*-phase change since* $n_0 < n_1$. *Interface centered at y = 16000. The transmitted fields are in phase but with amplitudes in the ratio of* $n_1 / n_0 = 3$, *and with pulse width reduced by* $n_1 / n_0$ *since the speed of the transmitted pulse is reduced by this factor. This is readily seen on comparing (a) t = 40k, and (b) t = 60k. Note the ratio of the fields* $|B_x / E_z| = n$, *where n is the refractive index of that medium.*

It is instructive to now consider the effect of this pulse propagating from high to low refractive index, rather than from low to high refractive index as in Figs. 11-13. Since the pulse speed in $n_0 = 3$ is a factor of 3 slower when compared with the speed in the low refractive index medium, the corresponding time outputs are a factor of 3 greater:

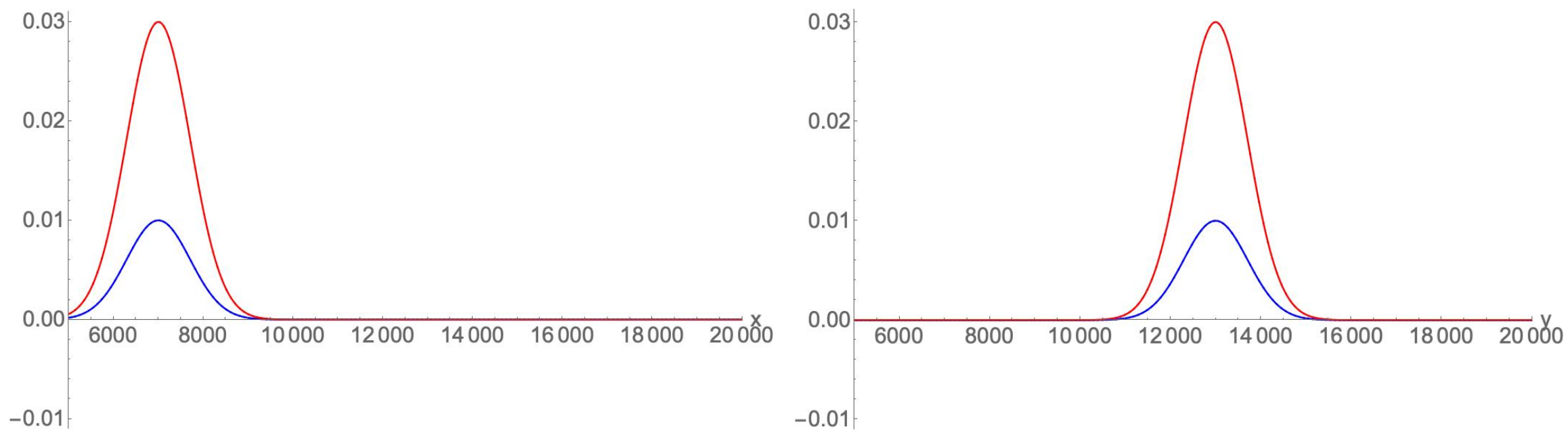

(a) t = 0 (b) t = 60 k

*Fig. 14 The propagation of a pulse from* $n_0 = 3$ towards $n_1 = 1$. *Note that* $B_x / E_z = n_0 = 3$ *for y < 16000.*

At this interface, it is now the magnetic field components $B_x$ that undergoes a $\pi$-phase change while the electric field component $E_z$ does not (as shown in Fig. 15):

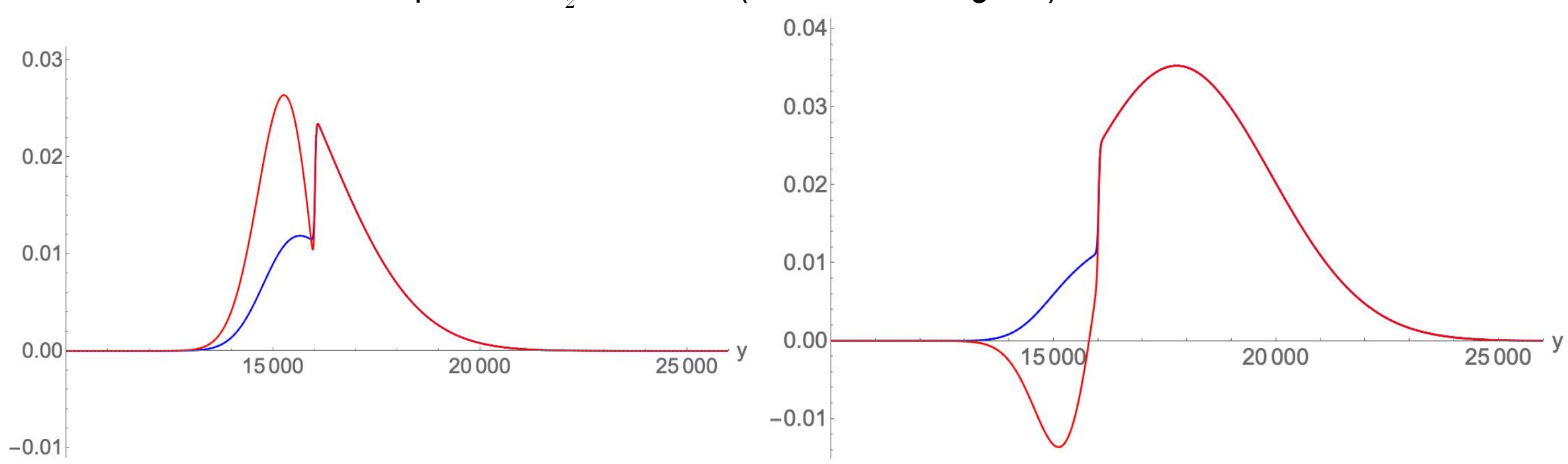

(a) t = 84 k (b) t = 96 k

*Fig. 15 The reflected and transmitted pulse for propagation from large to smaller refractive index.*
*(a) t = 84k : here the pulse is straddling the interface between the two dielectric media. Note that for y > 16000 the* $E_z(blue)$, $B_x(red)$ *profiles overlay each other;*
*(b) t = 96k : for y < 16000 there is a* $\pi$*-phase change in the reflected magnetic component* $B_x(red)$. *The axes had to be shifted because of the larger width of the transmitted pulse.*

At t = 120 K, Fig. 16 shows that the transmitted pulse has its width and speed of propagation increased by $n_1 / n_0$.

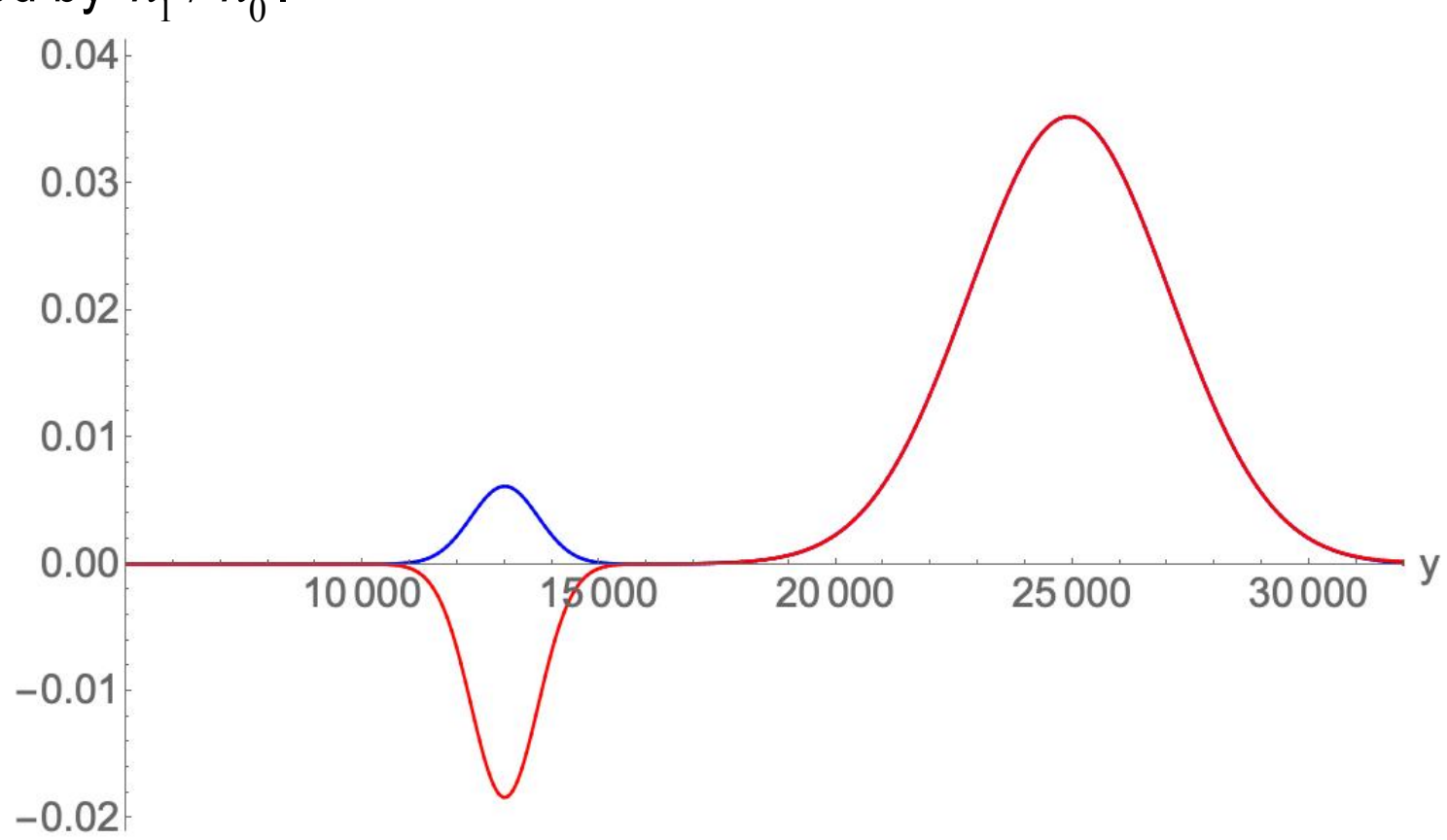

*Fig. 16 The reflected and transmitted pulse for propagation from* $n_0 \rightarrow n_1$ with $n_1 / n_0 = 3$ at t = 120k. *For the reflected pulse, there is no phase change in* $E_z(blue)$*, but a* $\pi$*-phase change in* $B_x(red)$*. The speed of the transmitted pulse is a factor of* $n_1 / n_0$ *greater than the incident (or reflected) pulse. The boundary between the media is at y = 16000. For the transmitted pulse, the* $E_z(blue)$, $B_x(red)$ *fields are equal and so overlay each other.*

Finally we show the reflection/transmission of a Gaussian wave packet as it propagates from a low-to-high refractive medium : $n_0 = 1$ to $n_1 = 3$. In the vacuum region, $n_0 = 1$ , $B_x = E_z$ so that these fields again overlay each other, Fig. 17. This overlay continues throughout the Gaussian packet's evolution through the vacuum region y < 16000 as can be seen in Fig. 17(b).

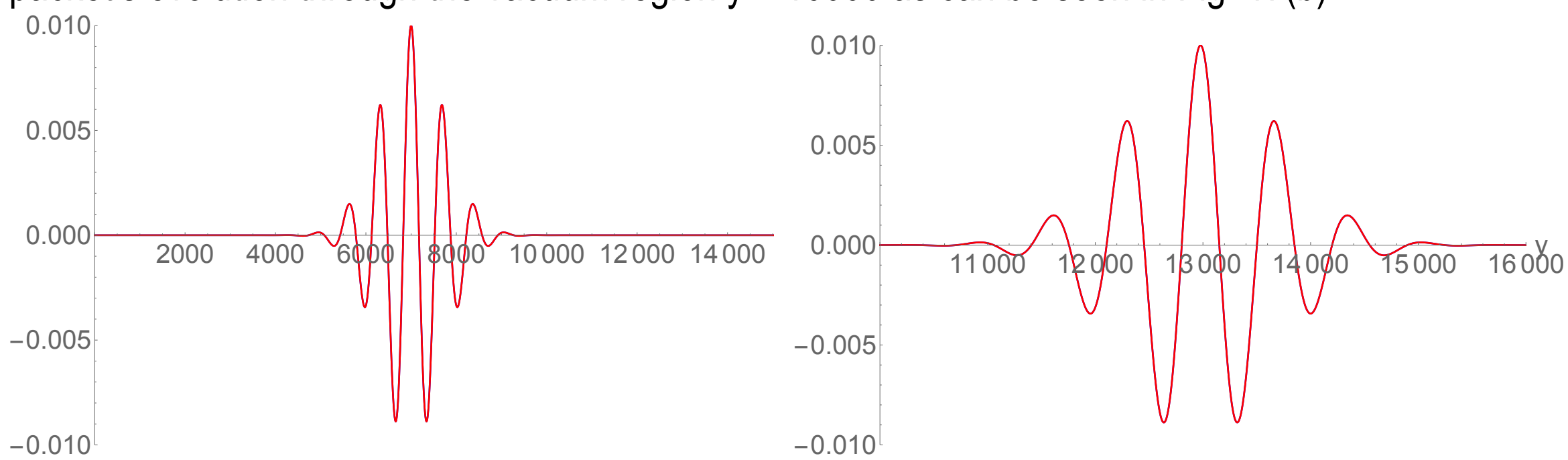

*(a) t = 0 initial profiles* *(b) The Gaussian wave packet at t = 20K*

*Fig. 17 (a) The initial Gaussian wave packet as it propagates from a vacuum* $n_0 = 1$ *to a higher refractive region for y > 16000 with refractive index* $n_1 = 3$*. Initially* $E_z = B_x$ *so the profiles overlay each other:* $E_z(blue)$, $B_x(red)$*. (b) The Gaussian wave packet at t = 20K as it approaches the boundary for the higher refractive index medium.*

In Fig. 18 the Gaussian packet, at t = 28k, is straddling the two dielectric media (interface at y = 16000). The $E_z(blue)$, $B_x(red)$ profiles no longer overlay near the interface and the wavelength of the packet oscillations in the denser medium are decreased over the vacuum wavelength.

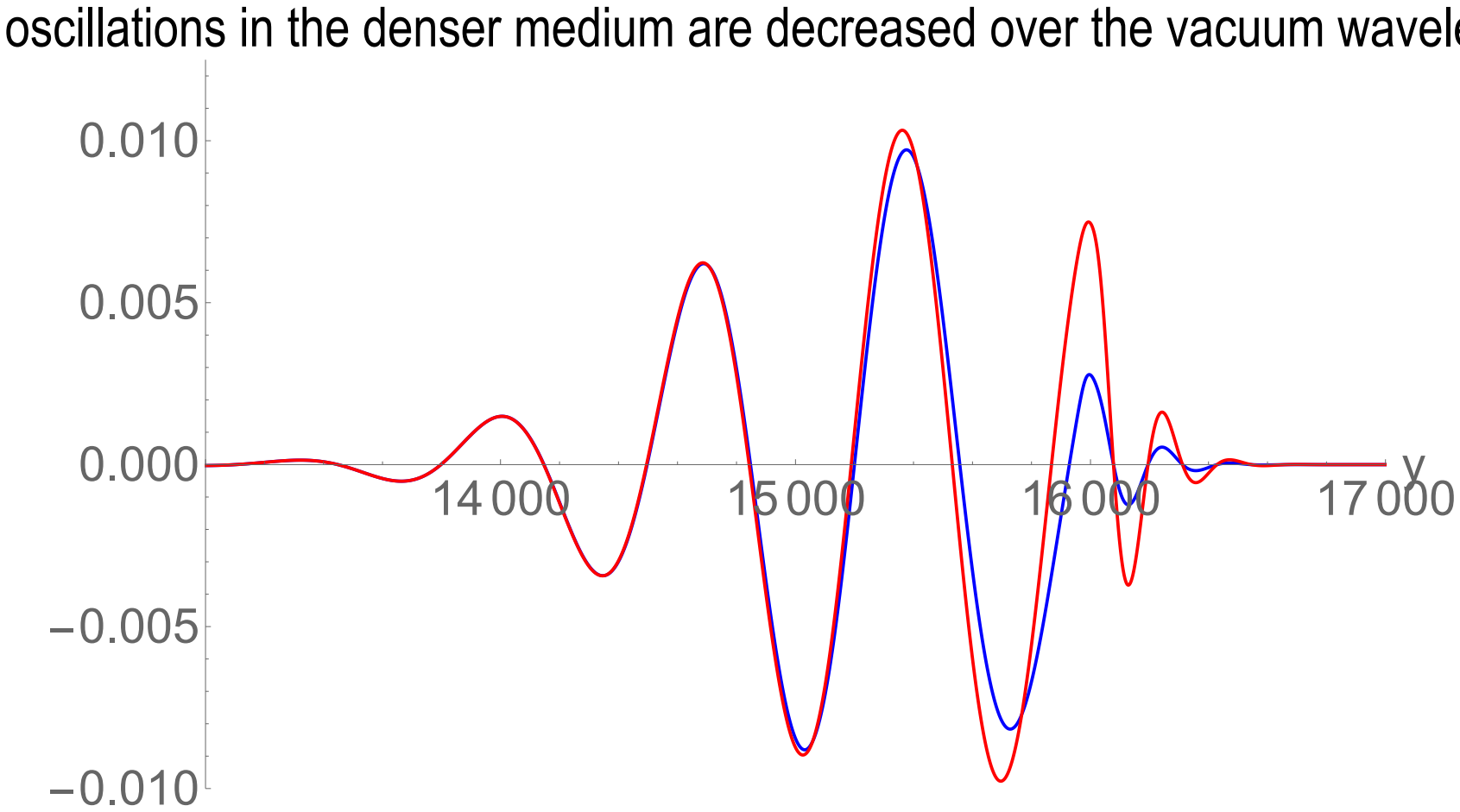

*Fig. 18 The Gaussian wave packet at t = 28K as it straddles the boundary between the two dielectrics :* $n_0 = 1$ for y < 16000 and $n_3 = 3$ for y > 16000.

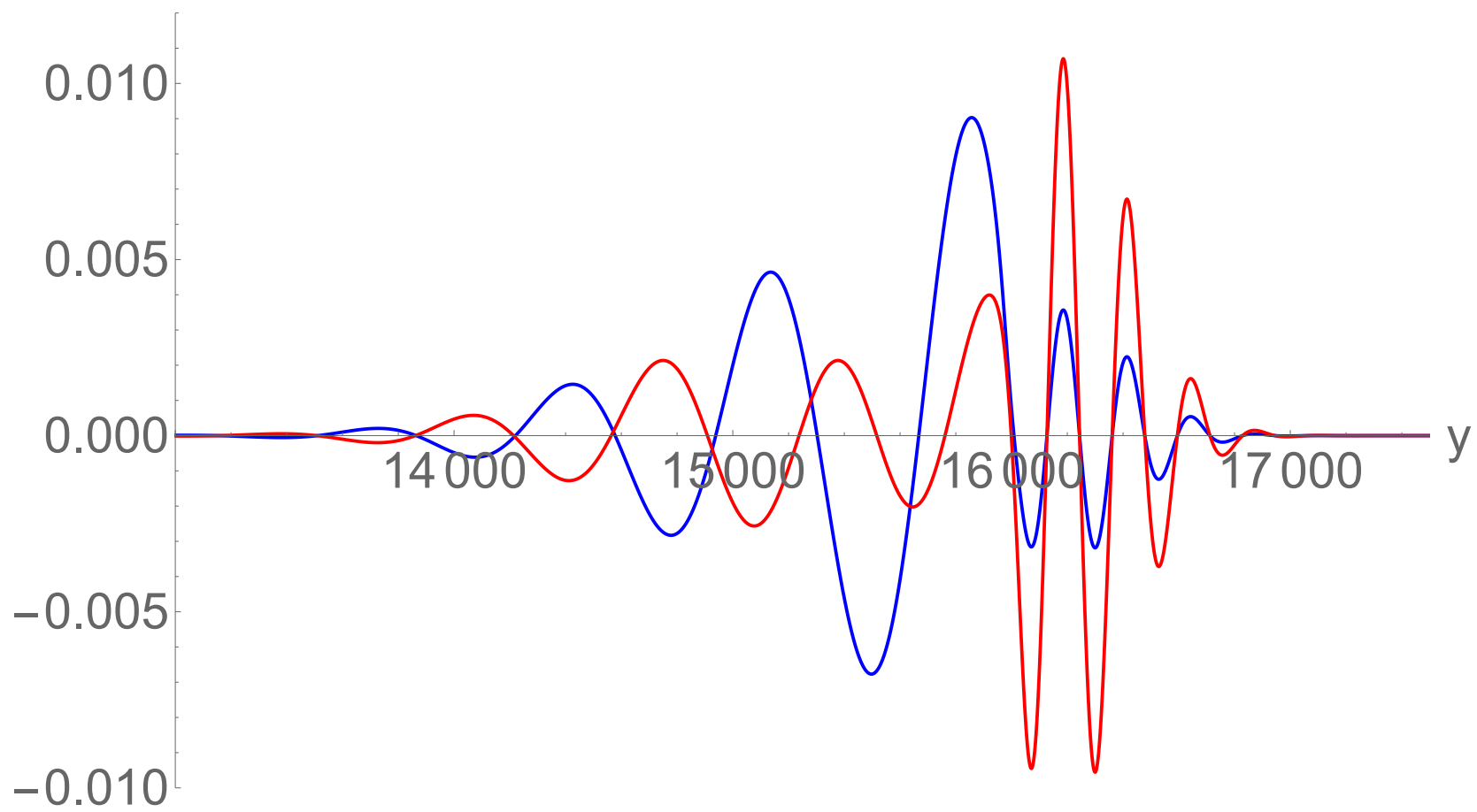

*Fig. 19 The start of the reflected and transmitted Gaussian wave packets in the vicinity of the dielectric interface at t = 32K. These profiles $E_z(blue)$, $B_x(red)$ no longer overlay each other in either dielectric region.*

The transient reflected and transmitted Gaussian wave packets are seen in Fig. 19, at t = 32K. Many features of the asymptotic profiles are becoming evident at this early stage : for the transmitted packet, the $E_z(blue)$, $B_x(red)$ profiles are in phase with $B_x = n_1 E_z$ and the wavelength of the transmitted packet is reduced by the factor $n_1 / n_0 = 3$. For the reflected packet one is clearly in a transient stage close to the interface but a phase difference of $\pi$ is evident for y < 15000. An asymptotic snapshot of the reflected and transmitted packets is seen in Fig. 20 at t = 40K.

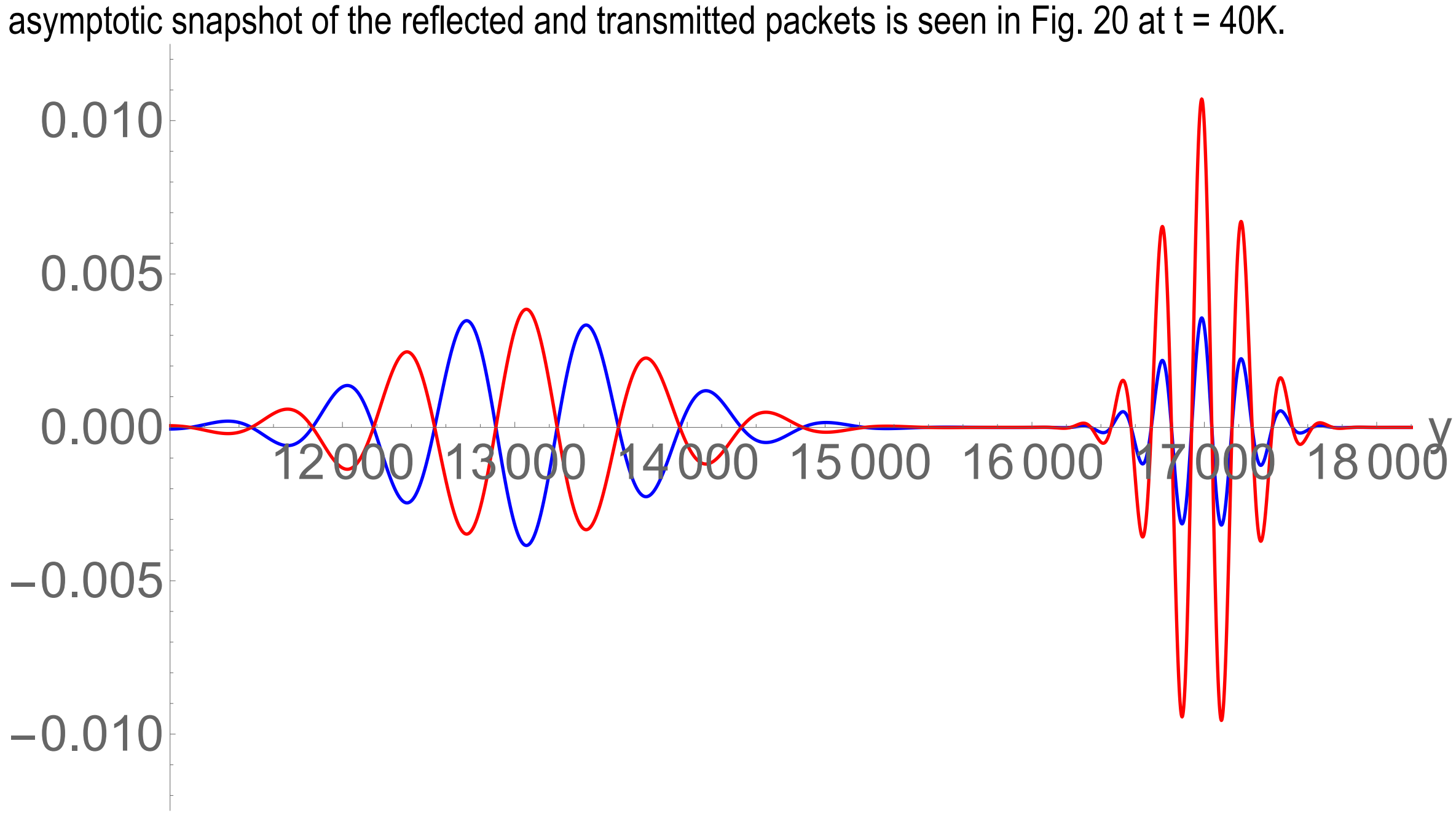

*Fig. 20 The asymptotic reflected (y < 16000) and transmitted (y > 16000) Gaussian wave packets arising from an incident Gaussian wave packet propagating towards the larger refractive medium at t = 40K. For the reflected packet, the $E_z(blue)$ field is $\pi$ out of phase with its incoming profile while the $B_x(red)$ field remains in phase. Also $-E_z = B_x$, For the transmitted packet, the field*

*components are in phase:* $E_z\left(blue\right)$, $B_x\left(red\right)$*, with wavelengths reduced by a factor of 3 and* $B_x = 3E_z$ *.*

## IV. SUMMARY and CONCLUSIONS

Utilizing the similarity of the spinor representation of the Dirac equation to the Maxwell equations, we have extended our studies in unitary QLA [8-23]. In particular, using the Pauli spin ½ - matrices, we have expressed Khan's Riemann-Silberstein representation of the Maxwell equations in a unitary spinor lattice representation. The QLA is readily determined for the 1D and 2D spatial dependence of the electromagnetic fields. For homogeneous media, the QLA requires only 4 spinor components per spatial lattice node, while for inhomogeneous media the two polarizations of the electromagnetic fields are coupled requiring the use of the two Riemann-Silberstein vectors and an 8-spinor. The QLA can be shown, using Mathematica, to be 2nd order accurate under diffusion ordering. To attain this ordering one must introduce a small parameter $\varepsilon$ into the unitary collision operators. Unlike in our earlier works of QLA for the Nonlinear Schrodinger equation and Bose-Einstein condensation for spinor fields, the introduction of the required small parameter can be accomplished by an appropriate scaling of the spinor order parameter wave function that appear in the nonlinear Bose-Bose interaction potential. For the Maxwell representation, this is not possible. By appropriately scaling the fields relative to the lattice spatial unit the QLA will still hold for sufficiently small $\varepsilon$ .

To benchmark the QLA we have considered two problems: (1) electromagnetic propagation in 2D homogeneous media, and (2) electromagnetic propagation in a 1D inhomogeneous media. In 2D homogeneous media, we have tested propagation in the x-direction and y-direction separately. This was done since the Pauli spin ½-matrices $\sigma_x$ from $\sigma_y$ have very different properties, resulting in different unitary collision-streaming operators. In 1D inhomogeneous media, we studied the well known and analytically soluble (for a plane wave) problem of a 1D normally incident electromagnetic wave onto a dielectric slab and the resultant transmitted and reflected waves. We show that the analytical results for the electromagnetic fields are asymptotically recovered by our QLA simulations of a Gaussian pulse. Moreover we can simulate the effect of the pulse as it impacts and enters into the dielectric region. The correct phase change of $\pi$ in the electric field component is recovered when the wave is incident onto a higher refractive index medium, while it is the wave magnetic field that undergoes a phase change of $\pi$ for wave propagation from higher to lower refractive index.

It is interesting to compare our 1D QLA, which utilizes simple unitary collision and streaming operators based on the Pauli spin ½ matrices, with the unitary Jestadt algorithm [30]. Jestadt et. al. [30] use spin-1 operators since they consider only 3-component spinors. They then approximate the unitary time evolution operator using split-operator methods, just as we do. They do not specify their collide-stream operator sequence nor the refractive index of their dielectric medium. However they invoke the Baker-Campbell-Hausdorff expansion to approximate the exponential operator in the commutator of their stream-collide operators and their inhomogeneous medium operator. This commutator involves the second derivative on the refractive index $n''\left(y\right)$. In our QLA, it is the interleaving of the non-commuting unitary collide-stream operators that yields the Maxwell equations - if was had ignored the non-cummatative property of the collision and streaming operators then our sequence would simply result in the identity operator itself.

The vista for further applications is boundless as the field of electromagnetic wave propagation in different dielectric media, like a 3D magnetized plasma (plasma..) lies before us.

## V. ACKNOWLEDGMENTS

The 2D simulations were performed on the DoD HPC supercomputer. GV, in particular, thanks Jeffrey Yepez for many fruitful discussion on QLA. LV was partially supported by an AFRL STTR Phase I with Semicyber LLC contract number FA864919PA049. AKR was supported by DoE Grant Number DE-FG02-91ER-54109.